# The Tetragonal to Orthorhombic structural phase transition in multiband FeAs-based superconductors

Alessandro Ricci, Michela Fratini and Antonio Bianconi

*Department of Physics, Sapienza University of Rome, P. Aldo Moro 2, 00185 Roma, Italy*

E-mail: antonio.bianconi@roma1.infn.it

**Abstract**

We report the temperature dependent x-ray powder diffraction of the FeAs-based superconductors in the range between 300 K and 95 K. In the case of NdOFeAs we have detected the structural phase transition from the tetragonal phase, with P4/nmm space group, to the orthorhombic phase, with Cmma space group, over a broad temperature range from 150 K to 120 K, centered at $T_0$ ~137 K. This transition is reduced, by about ~30K, by the internal chemical pressure going from LaOFeAs to NdOFeAs. On the contrary the superconducting critical temperature increases from 27 K to 51 K going from LaOFeAs to NdOFeAs doped samples. The FeAs layers in all undoped "1111" and "122" systems suffer a tensile misfit strain. The tensile misfit strain is reduced in "1111" and in "122" samples and at optimum doping the misfit strain is close to zero. This result shows that the normal striped orthorhombic Cmma phase competes with the superconducting tetragonal phase. In the orthorhombic clusters the charges can move only along the stripes in the b direction and are localized by the magnetic interaction.

## 1. Introduction

Recent experimental results in the high $T_c$ superconductors field have added to the known high $T_c$ superconductors [1]: the layered FeAs-based superconductors [2-19].
In this work we will show that the phase diagram of FeAs based superconductors shows a mesoscopic phase separation (20-22) in a region of the 3D phase diagram determined by the charge density, and the misfit strain of the superconducting layerrs separated by spacer layers (1, 23-27)
There are different types of layered FeAs-based superconductors that have an analogous structure:
a) Doped quaternary rare-earth iron oxypnictides, ROFePn (R = rare-earth metal and Pn = pnicogen

= O) (R=La, Pr, Nd, Ce, Sm…) made of FeAs layers intercalated by RO oxide layers. These "1111" systems at room temperature have a tetragonal (space group P4/nmm) structure. It is critical to the high Tc superconductivity (55 K is the maximum $T_c$) the F substitution for oxygen (15-20 atm%, called electron doping (n-type) of the formal [FeAs]- 1 layers; or $Sr^{2+}$ for $R^{3+}$ doping (4-13 atm%, called hole doping (p-type); or the introduction of oxygen defects.

b) Doped alkaline earth iron arsenides, $AeFe_2As_2$ (Ae=Sr,Ba), made of $[Fe_2As_2]$–2 layers separated by simple Ae-layers have a tetragonal $ThCr_2Si_2$-type, space group I4/mmm called "122". They become superconductors (38 K maximum $T_c$) with appropriate substitution of bivalent Ae cations with monovalent alkali metals, K, Cs … For example the K for Sr substitution of 45 atm% in $Sr_{1-x}K_xFe_2As_2$ gives the maximum $T_c$ [28-34].

c) Undoped compounds, $KFe_2As_2$ and $CsFe_2As_2$, made of $[Fe_2As_2]$–1 layers separated by monovalent ions are superconducting, with low $T_c$'s of 3.8 K and 2.6 K.

d) Undoped LiFeAs made of [FeAs]–1 layers is a superconductor with $T_c$ = 18 K [35].

e) Undoped non-superconducting $AeFe_2As_2$ (Ae=Ca,Sr,Ba) compounds, made of $[Fe_2As_2]$–1 layers, become superconductors under pressure. [36-38].

We have measured the misfit strain of the undoped parent compounds of FeAs-based superconductors RFeAsO systems by powder X-ray diffraction. The ROFeAs (R=La, Pr, Nd and Sm) powder samples have been synthesized in Bejing as described elsewhere [3-5]. The X-ray diffraction patterns were recorded at the x-ray Diffraction beam-line (XRD1) at the Elettra synchrotron radiation facility in Trieste. The lattice parameters of the 122" systems are taken for the literature. It is clear from the data that The FeAs layers in all undoped "1111" and "122" systems suffer a tensile misfit strain. The tensile misfit strain is reduced in "1111" and in "122" samples at optimum doping the misfit strain is close to zero [39]. A large tensile misfit strain promotes the low temperature charge and spin ordering phase that competes with superconductivity, and high Tc superconductivity prevails where the misfit strain goes to zero.

In this work we have investigated in particular the stoichiometric system NdOFeAs.

2. Results

In the range between 300 and 200 K the XRD diffraction pattern of the NdOFeAs shows the typical tetragonal structure with *P4/nmm* space group. By decreasing the temperature below 200–150 K the line shows an increasing broadening that increases rapidly in the range between 150 and 137 K. The data are fitted with the *Cmma* space group below 137 K, where we can clearly see the splitting of this line into the two lines of the *Cmma* space group that are indexed as 040 and 400. The results show that the tetragonal to orthorhombic phase transition, centered at 137 K, is a 30 K wide transition extending from 150 to 120 K, and the 040 and 400 lines are well resolved with our experimental resolution only below 137 K.

It is known that tuning the chemical potential at an electronic topological transition (ETT) the electron gas shows a 2.5 Lifshitz electronic topological transition; the compressibility of the electron gas becomes negative, therefore the system has the tendency toward a first order electronic phase separation. The electronic instabilities at the ETT's have been widely studied in the case of one-dimensional (1D) and two dimensional (2D) single band systems with the formation of 1D CDW and 2D CDW insulating phases respectively. All layered undoped parent compounds of the FeAs-based superconductors are multiband systems. In fact all undoped parent compounds show a similar tetragonal to orthorhombic transition occurring at low temperature $T_s$ detected by high resolution x-ray diffraction. We show in Fig. 1 the splitting of the "a" axis in stoichiometric ROFeAs "1111" systems at the structural transition from tetragonal (space group P4/nmm) to orthorhombic space group (Cmma) at low temperature, observed in the systems with R= La, Nd, Sm in agreement with previous works [3-10]. In Fig. 1 we report the XRD results for $AFe_2As_2$ "122" systems that show a similar structural transition from tetragonal $ThCr_2Si_2$-type, space group I4/mmm, to orthorhombic Fmmm space group [11-17]. In the orthorhombic phase a static stripe magnetic phase has been found. The structural transition takes place in a range of about 2 K in "122" systems, and it has been interpreted as first order transition [15-17] since it shows hysteretic behaviour. The structural transition in the "1111" systems shows a continuous character over a large temperature range above and below the critical temperature [5,6]. There is a strong coupling between magnetic and structural order parameters [18,19]. The spin ordering below the critical temperature $T_s$ is driven by the low temperature orthorhombic phase and it shows a striped phase with the antiferromagnetic coupling in the direction of the long Fe-Fe bond (the orthorhombic $a_o$ axis) and ferromagnetic coupling in the direction of the short Fe-Fe bond (the orthorhombic $b_o$ axis). The results in Fig. 1 clearly show that the critical temperature $T_c$ of the structural phase transition decreases with decreasing the tensile strain due to lattice misfit. It is possible to see that the $BaFe_2As_2$ case shows an anomalous behaviour.

The superconducting phase is observed to emerge from the non-superconducting magnetically ordered phase through appropriate doping of the charge reservoir spacer blocks. The FeAs-based materials are quite different from cuprates since the parent compounds are metallic systems and not Mott insulators. There is on the contrary a strong analogy with the high-$T_c$ cuprates if one assumes that the parent compound of all cuprates superconductors is the striped phase, at 1/8 doping and 7% misfit strain. In fact a few number of authors [2,20-27] have proposed that the relevant quantum critical point for high $T_c$ superconductivity in cuprates is where the superconducting phase competes with the striped phase, at 1/8 doping and 7% misfit strain.

The pressure-induced superconductivity in the non-superconducting compounds $AeFe_2As_2$ (Ae=Ca,Sr,Ba) indicates the role of the lattice in tuning the chemical potential. Therefore the high $T_c$ phase can be reached by varying the lattice parameters (modified by the external pressure or internal pressure) and the carrier densities in the $Fe_2As_2$ layers.

The pressure experiments in $K_{1-x}Ba_xFe_2As_2$ [2] and $K_{1-x}Sr_xFe_2As_2$ [28] show that the critical temperature is a function of both lattice parameters and charge density in the active FeAs layers and the maximum $T_c$ occurs along a line of points of charge density and pressure. There is now a large agreement that by using pressure, internal pressure (as it is shown in Fig. 1) and doping it is possible to decrease the temperature $T_s$ of the structural and magnetic phase transition toward zero. The system shows a mesoscopic phase separation (MePhS) of orthorhombic striped magnetic clusters and tetragonal superconducting clusters in the proximity of the quantum critical point for the structural phase transition. We show a pictorial view of this MePhS in Fig. 2 in a high-doping regime, where the average structure is the tetragonal lattice. In the orthorhombic clusters the charges can move only along the stripes in the b direction and are localized by the magnetic interaction in the direction. Therefore the first superconducting regime can be called a case of nematic electronic phase of itinerant fluctuating striped bubbles. Therefore in the proximity of the zero temperature transition from the average orthorhombic phase to the tetragonal phase (a quantum phase transition driven by charge density, chemical pressure or pressure) there should be a Fermi surface that fluctuates in space and time between a 2D topology in the tetragonal clusters and a 1D topology in the orthorhombic clusters.

In conlcusion we have shown that the FeAs based high Tc superconductors are expected to show a mesoscopic phase separation in the proximity of a first order phase transition [39], Therefore there is a similarity with cuprate superconductors since in both systems the superconducting phase merges in a phase of fluctuating bubbles of stripes, called superstripes [40], and the lattice charge instability can be manipulated by external fields [41,42].

**Acknowledgments:** We thank the staff of the XRD beam line of Elettra synchrotron radiation facility in Trieste and Naurang L. Saini and Nicola Poccia for help and discussions. We acknowledge financial support from European STREP project 517039 "Controlling Mesoscopic Phase Separation" (COMEPHS) (2005).

A. Ricci et al. arXiv:0812.0850

**Figure Captions:**

**Figure:1.** The structural parameter $a_0$, $b_0$ of the orthorhombic structure and $a_T\sqrt{2}$ of the tetragonal structure for the stoichiometric undoped parent compounds of the FeAs-based superconductors as a function of temperature, showing the structural phase transition from the high temperature tetragonal phase to the low temperature orthorhombic phase.

**Figure:2.**
a) a pictorial view of the orthorhombic phase.
b) a pictorial view of the fluctuating mesoscopic phase separation regime (MePhS) in the tetragonal phase, in the proximity of the structural phase transition from tetragonal (I4/mmm) to orthorhombic (Fmmm) structure. Where fluctuating nanoscale bubbles of striped magnetic matter (triangles) with quasi 1D Fermi surface coexists with the matrix of metallic phase (filled circles) with a 2D Fermi surface.
c) a pictorial view of tetragonal phase.

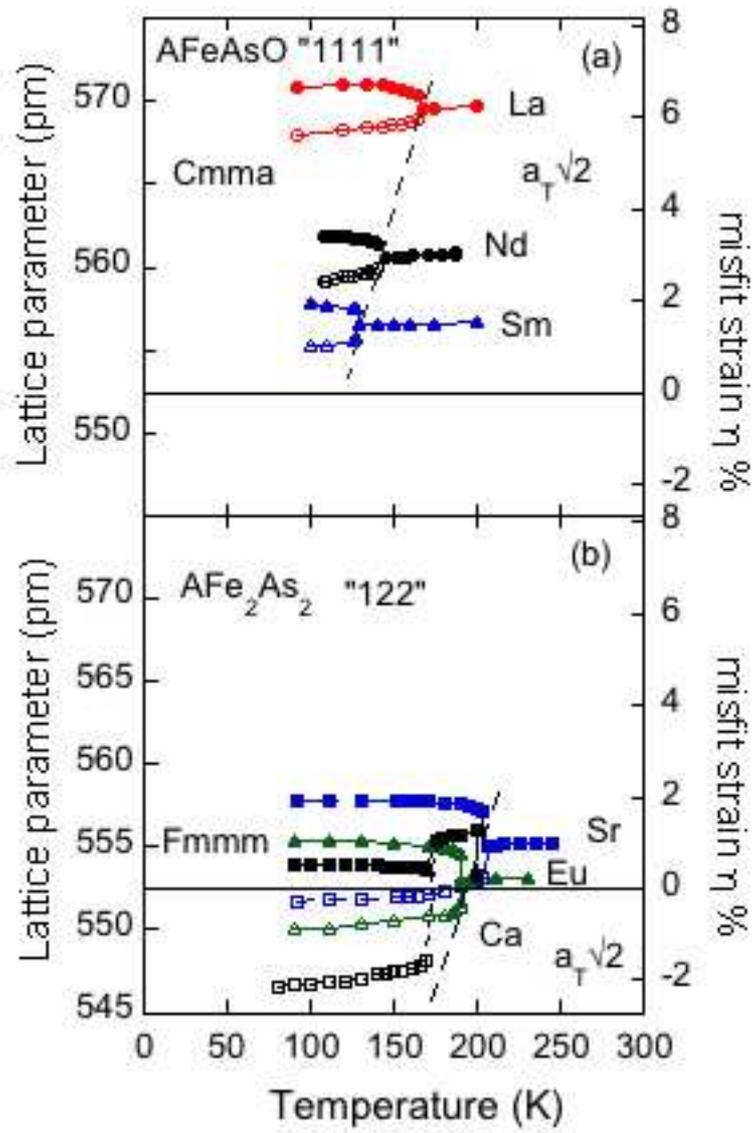

**Figure 1**

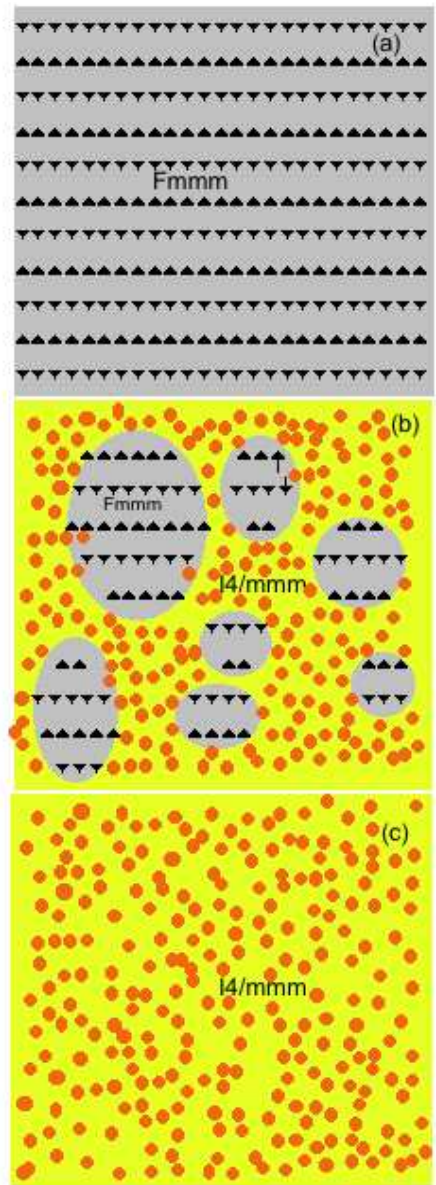

**Figure 2**